\documentclass[prl,showpacs,showkeys,preprint,floatfix]{revtex4}

\usepackage{graphicx}
\usepackage{units}
\usepackage{xspace}
\usepackage[american]{babel}
\usepackage{color}

\begin{document}


\title{Demagnetization cooling of a gas}

\author{M.\ Fattori}
\email{m.fattori@physik.uni-stuttgart.de}
\author{T.\ Koch}
\author{S.\ Goetz}
\author{A.\ Griesmaier}
\author{S.\ Hensler}
\author{J.\ Stuhler}
\author{T.\ Pfau}

\affiliation{E-mail:m.fattori@physik.uni-stuttgart.de Affiliation:
5.\ Physikalisches Institut, Universit\"at Stuttgart,
Pfaffenwaldring 57, 70550 Stuttgart, Germany}



\begin{abstract}
We demonstrate demagnetization cooling of a gas of ultracold
$^{52}$Cr atoms. Demagnetization is driven by inelastic dipolar
collisions which couple the motional degrees of freedom to the
spin degree. By that kinetic energy is converted into magnetic
work with a consequent temperature reduction of the gas. Optical
pumping is used to magnetize the system and drive continuous
demagnetization cooling. Applying this technique, we can increase
the phase space density of our sample by one order of magnitude,
with nearly no atom loss. This method can be in principle extended
to every dipolar system and could be used to achieve quantum
degeneracy via optical means.

\end{abstract}

\maketitle

Adiabatic demagnetization of a paramagnetic salt is the oldest
method for reaching temperatures in solids significantly below 1 K
\cite{Lounasmaa}. Realized for the first time in the 30's
\cite{DeHaas:1933}, nowadays it is still used for its simplicity
and flexibility. Based on the same principle of operation, nuclear
demagnetization refrigerators make use of nuclear instead of
electronic magnetic dipole moments \cite{Kurti:1956}. Crucial for
studies on magnetic phases in solids, they allow to cool the
nuclear spin system well below 1 $\mu$K
\cite{Oja:1997,Touriniemi:2000}. Also known as magnetocaloric
effect \cite{Pecharsky:1997}, demagnetization cooling works in
paramagnetic materials. They are constituted by particles with a
total angular momentum quantum number J and with a permanent
magnetic dipole g$\mu$(J(J+1))$^{1/2}$ where $\mu$ is the unit of
magnetic moment ($\mu_{\mathrm{B}}$ for electrons and
$\mu_{\mathrm{N}}$ for nuclei) and g$>$0 is the spectroscopic
splitting factor. If an external magnetic field is applied, such
dipoles will try to minimize their energy aligning to it and
generating a macroscopic magnetization M of the material. Quantum
mechanically, this can be easily explained looking at the
imbalance in the occupation probability $\propto$
exp(-E(m$_{\mathrm{J}})$/k$_\mathrm{B}$T) of every Zeeman state
with energy E(m$_{\mathrm{J}}$)=g$\mu$Bm$_{\mathrm{J}}$ and with
m$_{\mathrm{J}}$=-J, -J+1,....,J-1, J being the projection of the
dipole moment along the magnetic field. For an intense magnetic
field such that g$\mu$B$\gg$k$_{\mathrm{B}}$T the probability of
occupation of the state m$_{\mathrm{J}}$=-J is nearly one and M
saturates (see Fig. 1 a)). If after this isothermal magnetization
the sample is isolated and B is reduced, the sample demagnetizes
isoentropically and accordingly to the state equation
dQ=TdS=0=dE+pdV-BdM we get dE=BdM$<$0 since volume variations are
negligible in a solid \cite{Morrish:1983}. In other words, the
system has to do magnetic work to drive the demagnetization and
consequently its internal energy and therefore its temperature
decrease. This can be also understood from Fig. 1 b). When
k$_{\mathrm{B}}$T$\sim$g$\mu$B, Zeeman states with
m$_{\mathrm{J}}>$-J can be occupied at the expense of the energy
of the external degrees of freedom of the particles in the solid,
with a net cooling effect. The cooling efficiency can be
understood introducing the concept of the phonon reservoir, which
includes the external degrees of freedom of the particles, and the
spin reservoir to describe the internal state m$_{\mathrm{J}}$.
For high magnetic field, the spin degree of freedom is frozen and
the spin reservoir specific heat c$_{{\mathrm{s}}}$ is negligible.
The initial total energy E$_{\mathrm{i}}$ is then equal to the
phonon reservoir specific heat c$_{\mathrm{p}}$ times the initial
temperature T$_{\mathrm{i}}$. When k$_{\mathrm{B}}$T $\sim$
g$\mu$B, c$_{{\mathrm{s}}}$ becomes of the order of
k$_{\mathrm{B}}$ and if a coupling between the two reservoirs
exists, E$_{\mathrm{i}}$ has to redistribute over a system with a
total specific heat c$_{\mathrm{p}}$ + c$_{\mathrm{s}}$. The final
temperature is then
T$_{\mathrm{f}}$=T$_{\mathrm{i}}$(c$_{\mathrm{p}}$/(c$_{\mathrm{p}}$
+ c$_{\mathrm{s}}$)). Since in solids c$_{\mathrm{p}}$ $\ll$
c$_{\mathrm{s}}$, demagnetization can cool the sample by several
orders of magnitude.

\begin{figure}[hhh!]
 \includegraphics[width=0.5\textwidth]{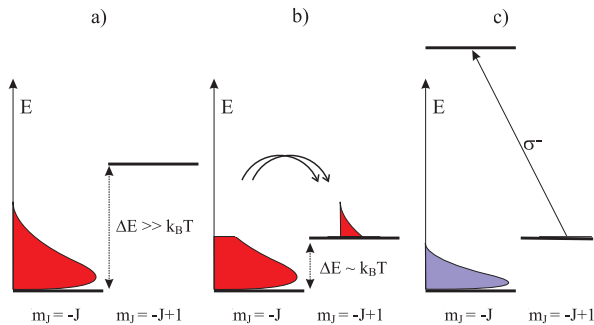}
  \caption{\label{figure1}Description of demagnetization cooling. a) Initially the particles are polarized
   in the Zeeman state m$_{\mathrm{J}}=-{\mathrm{J}}$ at high magnetic field B. The mean energy
  k$_{\mathrm{B}}$T is well below the Zeeman energy separation $\Delta $E=g$\mu_{\mathrm{B}}$B.
  b) Reducing the external magnetic field until k$_{\mathrm{B}}$T $\approx$ g$\mu_{\mathrm{B}}$B allows demagnetization
  of the system. Part of the energy of the sample is transferred into Zeeman energy with a net cooling effect. c) In an atomic gas it is
  possible to use optical pumping to polarize the atoms back to m$_{\mathrm{J}}=-{\mathrm{J}}$ leaving the temperature
  nearly unchanged. The system is ready for another cooling cycle.}
\end{figure}

Despite suggested by Kastler \cite{Kastler:1950} 25 years before
the first proposal on laser cooling \cite{Haensch:75},
demagnetization processes so far have not been implemented in
cooling techniques for gases. This is mainly due to the very weak
coupling between the spin and the external degrees of freedom
(phonon) reservoir in a system where the density is much smaller
than in a solid. Very recently, our group has revisited this old
idea and quantitatively analyzed its feasibility in the cooling of
atoms that show large inelastic relaxation rates between magnetic
substates \cite{Hensler:2005}. During such relaxation, the sum of
the quantum numbers m$_{\mathrm{J}1}$ and m$_{\mathrm{J}2}$ of two
colliding atoms is not conserved and the demagnetization of the
sample is allowed. Magnetic dipole-dipole interaction can induce
inelastic relaxations with a rate that drastically increase with
the atomic magnetic moments. In particular the cross section for
the inelastic single spin-flip (only one atom changes its
m$_{\mathrm{J}}$ value by one) is proportional to
J$^3$\cite{Hensler:2003a}. In \cite{Hensler:2005}, particular
attention has been paid to chromium atomic gases
\cite{Griesmaier:2005a} because grounded Cr $^7$S$_3$ atoms
possess very large magnetic dipole moments of six Bohr magnetons.
This is due to the spin of six unpaired electrons in outer shells.
As a consequence, in chromium the cross section for inelastic
single spin flips is a factor of $\sim$200 larger than in alkali
atoms (S=J=1/2).

In this Article, we demonstrate for the first time demagnetization
cooling of a gas. $^{52}$Cr atoms in the $^7$S$_3$ ground state
are initially polarized in the lowest energy Zeeman state (J=3,
m$_{\mathrm{J}}$=-3) with a magnetic splitting
2$\mu_{\mathrm{B}}$B much larger than the temperature T of the
sample. Reducing the magnetic field to values such that
3/2k$_{\mathrm{B}}$T $\approx$ 2$\mu_{\mathrm{B}}$B (100
$\mu$K$\sim$ 1 Gauss), transitions to higher energetic Zeeman
substates, caused by dipolar relaxation collisions, cool the
sample. In fact, the colliding atoms slow down because part of
their kinetic energy is converted into Zeeman energy of the
internal state. Single or double spin flips are possible. The main
advantage of demagnetization cooling in a gas is that by using
optical pumping it is possible to polarize the sample back to
m$_{\mathrm{J}}$=-3, constantly cooling the spin reservoir (see
Fig. 1 c)).

Our atoms are stored in an optical dipole trap realized by a 1064
nm fiber laser. A horizontal 20 W beam focused to a 30 $\mu$m
waist generates a harmonic potential, independently of the
magnetic substate, with $\omega_y/2\pi=\omega_x/2\pi = 2$ kHz,
$\omega_z/2\pi = 20$ Hz and a depth of 200 $\mu$K. Our experiments
start with $10^6$ atoms polarized in m$_{\mathrm{J}}$=-3. The
details of the loading can be found in a previous work
\cite{Griesmaier:2005a}. The initial temperature of the sample
(temperatures are measured using absorption immaging and time of
flight techniques) is $\sim 19 \mu$K due to plane evaporation and
the external magnetic field is 1 Gauss. If we suddenly decrease
the magnetic field to 50 mG and let the system evolve, we observe
a reduction of the temperature on a timescale of few seconds (see
black squares in Fig. 2a)). After 5 seconds the equilibrium is
achieved and we measure a final temperature of 16 $\mu$K. We have
repeated the measurement keeping a 1 Gauss external magnetic field
(see red circle in Fig. 2a)). As expected no temperature reduction
has been detected.
\begin{figure}[hhh!]
 \includegraphics[width=0.4\textwidth]{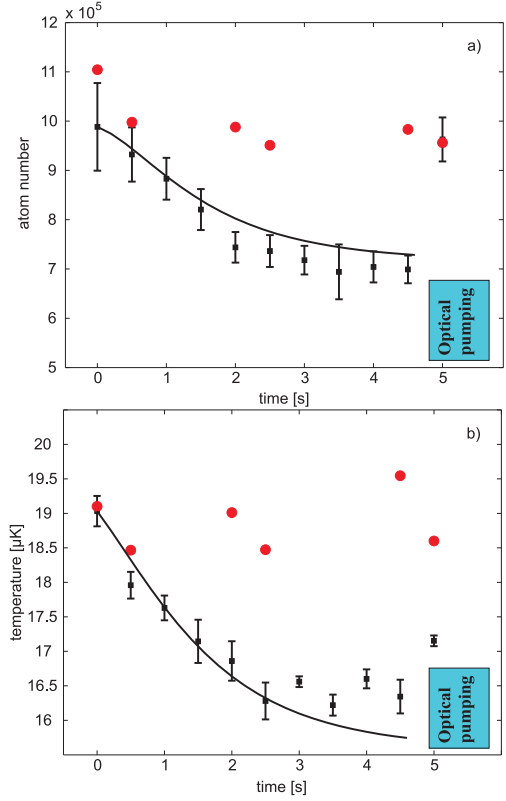}
  \caption{\label{figure2}Single step demagnetization. At time zero the external
  magnetic field is suddenly switched from 1 G to 50 mG. Temperature (in Fig. a)) and atom
  number (in Fig. b)) evolutions are represented by black squares. Every point is the result
  of three measurements average. Red circles show how the system evolves
  if the external magnetic field is kept at 1 G. Demagnetization results in temperature
  reduction and atomic depolarization, how our state selective measurement reveal.
  Switching on the optical pumping beam after 5 seconds slightly affects the temperature
  and pumps back all the atoms to the initial state. The black curve is the result of a theoretical calculation that
  takes into account demagnetization through all the seven magnetic sublevels. The input parameters
  are only the atom number, the initial temperature and the trap frequencies.}
\end{figure}
Since the energy of atoms is much larger than $\hbar
\omega_{\mathrm{i}}$ (i=x, y, z) the gas is classical and the
specific heat per particle of the external degrees of freedom
reservoir is 3k$_\mathrm{B}$. Considering that the spin reservoir
contribution is of the order of k$_B$ we can then explain why the
single step demagnetization can reduce the temperature by a factor
of $\sim$ 3/4. Another proof of the demagnetization of the system
comes from the atom number measurement. Our detection scheme makes
use of $\sigma^{-}$ light resonant with the $^7$S$_3$
m$_{\mathrm{J}} =-3$ - $^7$P$_4$ m$_{\mathrm{J}}' =-4$ transition.
Absorption signal depend on the atomic distribution over the
Zeeman substates and is stronger for atoms polarized in
m$_{\mathrm{J}}=-3$ (for m$_{\mathrm{J}}\ne -3$ we have a
different Zeeman detuning and a different coupling strength CG$^2$
to the m$_{\mathrm{J}}'=m_{\mathrm{J}}-1$). As we can see in Fig.
2b) demagnetization results in a reduction of the detected atom
signal. To exclude atom loss, we repolarize the sample with
optical pumping after 5 seconds and verify that the detection
signal is as large as at the beginning. Again red circles in Fig.
2b) show that this cooling step does not cause any extra loss
respect to the high magnetic field reference case. From the
initial slope of the atom number and temperature curves in Fig. 2)
it is possible to measure the depolarization rate at t=0
\begin{equation}\label{equation1}
\frac{dN_{-3}}{dt} = - \beta_{\mathrm{dr}}
\frac{N_{-3}^{2}}{\overline{V}}
\end{equation}
where $\overline{V}=(\sqrt{4\pi
\mathrm{k}_{\mathrm{B}}{\mathrm{T}/m}}$ $)^3/(\omega_x \omega_y
\omega_z)$ is the mean volume of the atomic cloud and
$\beta_{\mathrm{dr}} = \langle (\sigma_1 +
2\sigma_2)v_{\mathrm{rel}}\rangle_{\mathrm{therm}}$. Averaging
over the inter-particle velocities we consider that the atomic
depolarization can occur via single or double spin flip. The
measurement of $\beta_{\mathrm{dr}} \sim 10^{-13}$cm$^2$/s results
in good agreement with theoretical predictions based on
1$^{\mathrm{st}}$ order Born approximation for dipole-dipole
interaction \cite{Hensler:2003a}.

Performing optical pumping at the end of the demagnetization is
crucial for preparing the system for another cooling step with
smaller magnetic field and lower final temperature. This can be
accomplished using $\sigma^-$ polarized light on the
$^7$S$_3$-$^7$P$_3$ optical dipole transition ($\lambda = 427
$nm). In this way m$_{\mathrm{J}}$=-3 is a dark state and its
population is not affected by the pumping light. It is then
possible to substitute several cooling steps with a continuous
ramp of the magnetic field keeping on the optical pumping beam
(OPB) during all the sequence. It has been proved theoretically
that such strategy increases the cooling efficiency
\cite{Hensler:2005}.

In order to preserve the polarization of the OPB
($I_{\sigma^+}/I_{\sigma^-} \sim 1/1000$) we have to keep the
external magnetic field aligned to the OPB's propagation axis
(y-axis). In fact, minimizing the heating effect of the OPB on the
atoms we have been able to compensate residual magnetic fields
B$_x$ and B$_z$ down to a few mG.


The continuous cooling sequence starts with 10$^6$ atoms at the
temperature of 19 $\mu$K with B$_y=$ 250 mG. The OPB is red
detuned 40 $\Gamma$ ($\Gamma = 2\pi \times 5$ MHz) from resonance
and the total scattering rate is $\Gamma_{\mathrm{op}}$= 200
photons/s. The best ramp of B$_y$, optimized experimentally in
order to maximize the final phase space density, results in a
linear ramp down to 50 mG in 7 seconds. Temperature and atom
number evolution can be seen in Fig. 3). Red circles allow a
comparison to the high magnetic field case. The final temperature
is 11 $\mu$K. The atom loss is not related to the cooling
mechanisms but most probably to finite background gas pressure.
The cooling is insensitive to the detuning of the OPB both on the
red and the blue side of the resonance in the 2-40 $\Gamma$
interval. Lower temperatures have not been achieved reducing B$_y$
even further. In the next paragraph we analyze possible
limitations.

As far as we are performing optical pumping largely detuned from
resonance and the saturation parameter s is $\sim 10^{-5}$, the
atom light scattering is mainly coherent
\cite{Cohen:92coherent_scattering}. Considering only the first
order process, the light re-emitted by the atoms is blue shifted
with respect to the pumping light by an amount
2$\mu_{\mathrm{B}}$B. Due to the random direction of the emitted
photon every cooling cycle causes an extra kick to the atoms with
a total momentum $\hbar k$ where $k = 2\pi /\lambda$ is the light
wavevector.
\begin{figure}[hhh!]
 \includegraphics[width=0.4\textwidth]{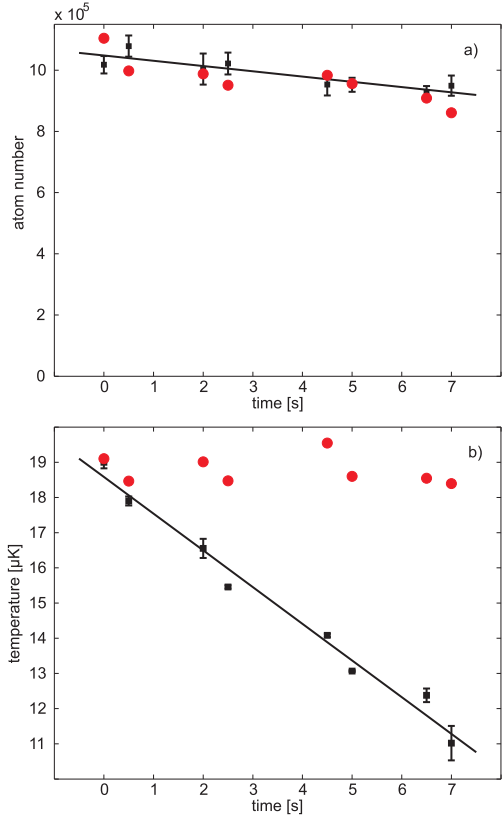}
  \caption{\label{figure2}Continuous demagnetization cooling. The external magnetic field is
  linearly ramped from 250 mG to 50 mG in 7 seconds. Black squares shows atom number (Fig. a)) and
  temperature (Fig, b)) evolution. Every point is the result of 3 measurements. Red circles describe the evolution for a
  constant external magnetic field of 1 G. Note that the atom number reduction is not due to the
  cooling but due to single particle losses probably associated with finite lifetime in the dipole trap.
  Lines fit the cooling measurements.}
\end{figure}
However, simulations show that the ultimate temperature limit is
slightly below the recoil temperature T$_{\mathrm{rec}}$=$1 \mu$K
\cite{Hensler:2005}. Reabsorption of the scattered photons could
be also a serious limitation. The on-resonance light absorption
cross section $\sigma = 6\pi/k^2$ holds the same even if the
fluorescence photon is far from resonance with the bare atom: in
fact the atoms are dressed by the optical pumping light and their
fluorescence frequency matches the dressed states' energy
difference \cite{Proceedings:1997}. Using the theoretical analysis
presented in \cite{Castin:1998}, which is valid for kr $\ge$1,
where r is the interparticle separation, we can calculate the
reabsorption probability. Note that the maximum density n $\sim
10^{13}$ atoms/cm$^3$ at the end of the cooling and the optical
pumping wavevector k fulfill the kr $\ge$1 condition. The
reabsorption probability is
\begin{equation}\label{equation1}
p \sim \frac{6\pi}{k^2}
\frac{\Gamma_{\mathrm{op}}}{\omega_{\mathrm{D}}} \frac{1}{4\pi
\langle r^2 \rangle} N
\end{equation}
The on-resonance cross section is reduced by a factor
$\Gamma_{\mathrm{op}}/\omega_{\mathrm{D}} \sim 10^{-4}$ where
$\omega_{\mathrm{D}} \sim 2 \times 10^{5}$ s$^{-1}$ is the Doppler
broadening at 10 $\mu$K. $\langle r^2 \rangle$ is the mean square
distance between two individual atoms in the trap ($\sim 3 \times
10^{-7}$ m$^2$) and N$\sim$10$^6$ the total number of trapped
atoms. Consequently, for our experimental parameters the
reabsorption probability is on the order of $p \sim 10^{-6}$.
Theoretical analysis of the cooling including recoil energy,
reabsorption, background gas and three body collisions predict a
final temperature of $\sim 1$ $\mu$K \cite{Hensler:2005}. We
therefore conclude that a heating mechanism is currently limiting
us, most probably coming from a non perfect control of the fields
B$_z$ and B$_x$. We observe in fact a drift in the currents that
optimize such transversal fields. Several checks are necessary
during one day measurement. This could be due to instability of
our current generators or to random magnetization of our steel
chamber during the switching of the magnetic trap we use to
prepare our atomic samples. External ac fields of the order of few
mG could also be limiting us.

The development of such cooling technique helps us on our route to
the chromium BEC. Our measurements show that the efficiency of
demagnetization cooling
$\chi=-\ln(\rho_{\mathrm{f}}/\rho_{\mathrm{i}})/ln($N$_{\mathrm{f}}/$N$_{\mathrm{i}})$
associated to the gain in phase space density over the loss in
atom number is $\sim 11$. So far, this is much better than the
optimum value achieved ($\sim 4$) using evaporative cooling. By
this our starting conditions in the optical dipole trap result
better, with nearly one order of magnitude higher phase space
density at the same atom number.


In the atom optics community, demagnetization cooling belongs to
the family of laser cooling techniques in external fields
\cite{Cirac:95, Cirac:96}. What is new is the mechanism that
performs the selection of higher energy atoms. In our case, we
make use of dipole-dipole inelastic collisions, while in other
experiments this is performed via optical Raman transitions
\cite{Lee:96, Kerman:2000a}. Such a technique, applied to Cs and
performed in a far-off-resonant lattice has lead to final phase
space density of 1/30 \cite{Han:2000a}. Higher values are
prevented by inelastic hyperfine changing collisions occurring
between two atoms at the same lattice site. Cr has no hyperfine
structure and this will not be the limiting effect in the
challenge of achieving Bose Einstein condensation via all-optical
means with demagnetization cooling. As shown in \cite{Wolf:2000a}
crucial for the suppression of reabsorption will be working in the
festina lente regime, where reabsorption is prevented if the
optical pumping rate $\Gamma_{\mathrm{op}}$ is smaller than the
trapping frequency. Such regime has been theoretically proved to
work if kr $\ge$1 \cite{Castin:1998} ( for higher densities
radiative interatomic collisions start to play an important role
\cite{Hijmans:96}). Considering our optical pumping light
wavelength and the mass of Cr we deduce that such condition is
fulfilled for typical condensation temperatures below 3 $\mu$K.
All optical BEC with demagnetization cooling is then in principle
possible with a more accurate control of the external magnetic
fields.

Demagnetization or the general depolarization cooling is an
important tool that can in principle be applied to other bosonic
and fermionic systems. Promising is its use in the cooling of
heteronuclear molecules with a permanent electric dipole moment.
Note that as dipole-dipole interaction is long range, in the
partial waves decomposition of the cross section all the orders
contribute, even in the limit of zero collision energy. Therefore,
also fermionic dipolar molecules and fermionic atoms can
demagnetize and thermalize via elastic dipole-dipole interaction.

\textbf{Acknowledgments}

We thank our atom optics group for encouragement and practical
help. This work was supported by the Alexander von Humboldt
Foundation and the German Science Foundation (DFG) (SPP1116 and
SFB/TR 21).

\textbf{Competing financial interests}

The authors declare that they have no competing financial
interests.


\end{document}